\documentclass[12pt,aps,pra,article]{revtex4-1}
\usepackage{amsmath}
\usepackage{graphicx}
\usepackage{subfig}
\usepackage{color}
\begin{document}

\title{Codimension 2 and 3 situations in a ring cavity with elliptically polarized electromagnetic waves}
\author{Daniel A. M\'{a}rtin and Miguel Hoyuelos}





\affiliation{Departamento de F\'{\i}sica, Facultad de Ciencias Exactas y
Naturales, Universidad Nacional de Mar del Plata and Instituto de
Investigaciones F\'{\i}sicas de Mar del Plata (Consejo Nacional de
Investigaciones Cient\'{\i}ficas y T\'{e}cnicas), Funes 3350, 7600 Mar del Plata, Argentina}

\begin{abstract}
We study pattern formation on the plane transverse to propagation
direction, in a ring cavity filled with a Kerr-like medium, subject
to an elliptically polarized incoming field, by means of two coupled
Lugiato-Lefever equations. We consider a wide range of possible
values for the coupling parameter between different polarizations,
$\bar{B}$, as may happen in composite materials. Positive and also
negative refraction index materials are considered. Examples of
marginal instability diagrams are shown. 
 It is shown that, within the model, instabilities cannot be of
codimension higher than 3. A method for finding parameters for which
codimension 2 or 3 takes place is given. The method allows us to
choose parameters for which unstable wavenumbers fulfill different
relations.  Numerical integration results where different
instabilities coexist and compete
 are  shown.
\end{abstract}


\maketitle


\section{Introduction}

Spatiotemporal patterns in non linear optical systems, along the
plane transverse to light propagation, have been widely studied both
theoretically and experimentally \cite{lugiatoRev,LibroST}.
Studies of optical patterns have some
common features with the analysis of pattern formation in other
physical systems, but there are also some specific aspects, such as
the role of diffraction and the vectorial degree of freedom associated with light
polarization.

  Patterns taking into account  the vectorial degree of freedom of incident fields were
analyzed in \cite{hoyuel1} for isotropic positive refractive index material (PRM) with third order
nonlinearities, i.e., a Kerr medium, and mainly for a specific value
of the nonlinear parameter ($\bar{B}=1.5$).
In composite materials an  enhancement of nonlinear polarizability
\cite{Yano}, and also a wider variety of nonlinear parameters
\cite{SipeyBoyd} may be expected.

An example of composite materials are
 negative refraction index materials (NRM);
they are materials with periodic inclusions 
which allow the experimental observation of novel optical properties, and for which several applications have been proposed \cite{rama}.
For standard PRM, there are arguments to neglect magnetic response,
but  these arguments do not
hold for NRM \cite{Merlin}.
 It has been also shown that an NRM can develop a macroscopic effective  nonlinear magnetic response \cite{zharov}. 
Negative diffraction is also expected in NRM,  but this property can also be obtained in regular, periodic refractive index materials \cite{Negdifrac,paperST2}. Soliton formation under zero or negative diffraction has already been studied \cite{paperST}.

 Here, we extend the study  of pattern formation in a ring cavity under arbitrary polarized fieds, so that it is valid for composite materials, 
either PRM or NRM, where other values of the nonlinear parameter (different form $\bar{B}=1.5$) may be expected, 
  and where  nonlinear magnetic response may or may not be present. 

We 
present a method for the analysis of eigenvalues that allows the
derivation of some exact and general results. The method allows us
to find parameters for  codimension 2 Turing-Turing (where two
different transverse wavenumbers destabilize simultaneously), Turing-Hopf and
codimension 3 Turing-Hopf-Turing instability. Also,  following our
analysis, it can be shown that codimension higher than 3 is not
possible within the model.

These results are then used in order to numerically integrate equations, and some results that are not found in codimension 1 situations are found. 
\section{The system}

\begin{figure}

\includegraphics[scale=0.3]{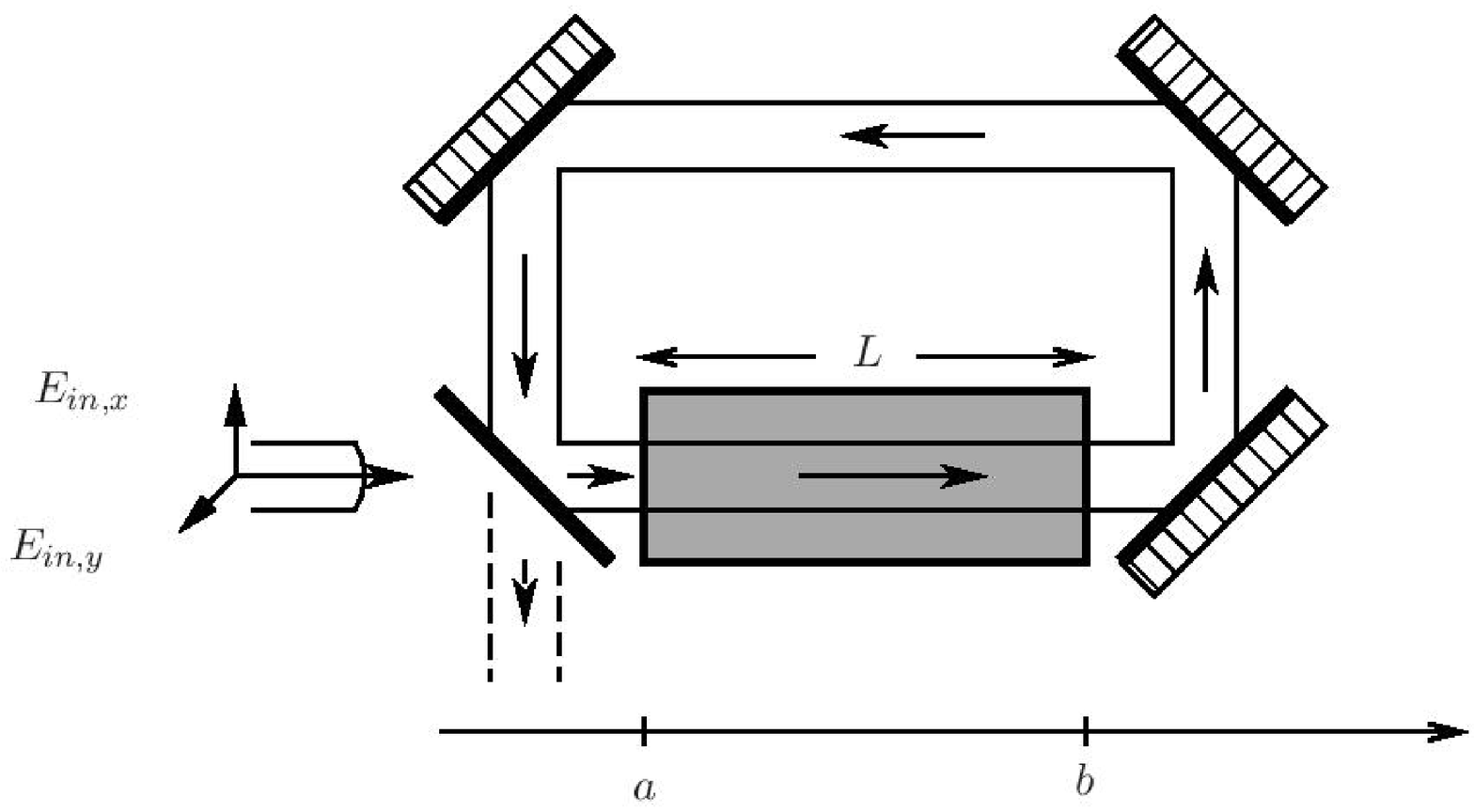}
\includegraphics[scale=0.3]{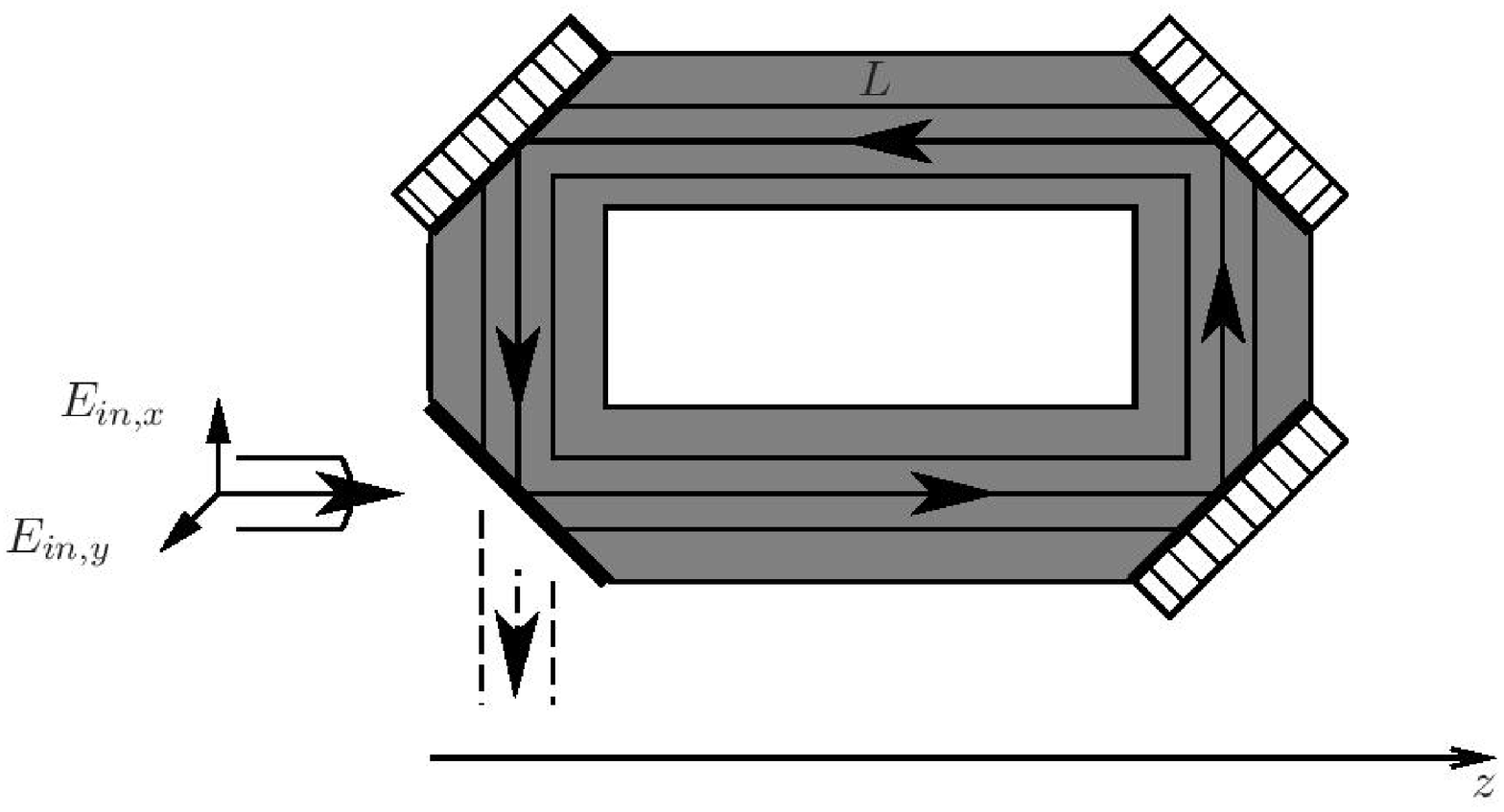}
\caption{Scheme of possible ring cavities. In the first one, the
nonlinear material has length $L$. In the other one, the material
fills the cavity and $L$ is the roundtrip length. Both are described
by the same equations.} \label{cav}
\end{figure}

The system under study is essentially the same as in  \cite{martin2} with the addition of the transverse spatial dependence. 
We consider a ring cavity with plane
mirrors filled  by an isotropic material or
metamaterial  with a third-order Kerr-like nonlinear response. Two possible  sketches of the system are
shown in Fig. \ref{cav}.

The field inside the cavity is described by a plane wave of
arbitrary polarization,  modulated by a slowly varying envelope. We
assume that the electric and magnetic fields are in the $x$-$y$
plane and the wave propagates in the z axis.   We study the cavity
close to resonance.

Based on the work by Zharov \emph{et al} \cite{zharov}, we allow the
material to have a nonlinear magnetization, which depends on the
magnetic field.

Light propagation in a Kerr-type PRM can be described by a nonlinear
Schr\"{o}dinger equation, 
and the same equation can be extended to NRM \cite{Tassin}. This
equation can be used to obtain the behavior inside a cavity.
 Taking into account the magnetic response, and applying the same process, we obtain
four  nonlinear Schr\"{o}dinger equations (two  for the envelopes of
electric fields and two for the envelopes of magnetic fields, defined in
the plane perpendicular to the $z$ axis). It  can be shown that
the magnetic field remains proportional to the electric field. So,
the system is well described knowing only the electric field.  The
procedure is analogous to that performed in \cite{nosotros}.

\section{Equations} 
\label{Equ}

After a change of variables,  two
coupled Lugiato-Lefever \cite{lugiato}  equations, describing
the left
and  right circularly polarized field amplitudes inside the cavity, can be obtained:
 \begin{eqnarray}\label{evol}
\frac{\partial A_\pm}{\partial t}= A_{in\pm} -(1+i \Theta) A_\pm + i h \nabla_\perp^2 A_\pm + \nonumber \\
i \alpha  \left[|A_\pm|^2
 \left(1-\frac{\bar{B}}{2}\right)+ |A_\mp|^2 \left(1+\frac{\bar{B}}{2}\right)\right] A_\pm ,
\end{eqnarray}
where all cuantities are adimensional, time and transverse coordinates have been normalized; $A_{\pm}$
are the normalized amplitudes of the electric field with circular
polarization (see \cite{martin2}), $\Theta$ is related to the cavity
detuning; $\alpha$ as the sign of $\chi^{(3) }_M  \eta^2 +
\chi^{(3)}_ E$, with $\chi^{(3) }_{E/M}$ being the transforms of
$xxxx$ component of the third order nonlinear electric and magnetic
tensors evaluated at $(\omega_0,\omega_0,-\omega_0)$ and $\eta$ the
inverse of the impedance. The  transverse Laplacian,
$\nabla_\perp^2$, refers to the second derivatives with respect to
the adimensional coordinates $x'=x/l$ and $y'=y/l$, where $l$ is a
characteristic distance (see \cite{nosotros}), $h=\pm 1$ is the
sign of the diffraction effects, which, in our model, is the same as
the sign of the refractive index. Notice, however, that negative  
 refractive materials are not necessary for negative 
diffraction: negative (and zero) diffraction resonators can be obtained in negative
 (or zero) effective length cavities built by means of curved mirrors,
 see \cite[Chapter 6]{LibroST}, or by means of a spatially modulated refractive index material, see \cite{paperST2}.   

The nonlinear parameter $\bar{B}$ is related to
%
%
components of the polarization and magnetization
tensors that measure the coupling between orthogonal
polarization
%
(the nonlinear parameter for the electric case is
defined in \cite{boyd}, and the generalization for magnetic
nonlinearities is explained in \cite{martin2}). 
Theoretical models predict  
(see \cite[p. 227]{boyd}): $\bar{B}
= 3/2$ in materials where nonlinearity is due to molecular orientation effects; and $\bar{B} = 2/3$ for electronic response far from resonance. 
However, in experiments  with SiO$_2$ subject to relatively long pulses, a value of  $\bar{B}$  as  low as  0.244 was measured \cite{Bugin}, which was explained as the effect of the competition between electronic and nuclear nonlinearities.
Also, the inclusion of small spherical particles inside a material, one or both having third order nonlinear response,  would result in a material where
 nonlinear effects might be greatly enhanced, and $\bar{B}$ may take a large range of values \cite{SipeyBoyd}. The inclusion of magnetic nonlinear effects in the analysis
gives more flexibility to the possible values for $\bar{B}$.


In general, we have $\alpha=1$. The less frequent case of
$\alpha=-1$ is equivalent to $\alpha=1$, and $\Theta$ and $h$ with
reversed signs, as can be seen by taking the complex conjugate of
Eq.\ \eqref{evol}. When we have only electric
nonlinearities, the case $\alpha=-1$ corresponds to a
self-defocusing material.  In the following, we assume $\alpha=1$.
We also assume that $|\Theta|<\sqrt{3}$; within this choice, bistable symmetric solutions are not present and changes in
 $\Theta$ do not modify qualitatively the results.

In the rest of our work, we
will limit our numerical results to the case $0 \leq \bar{B} \leq 2
$, 
and where $\chi_M^{(3)}$ has the same sign as $\chi_E^{(3)}$. 


Eq. (\ref{evol}) is robust in the sense that  not exactly matching
impedances can be allowed, and small dissipation can be taken into
account if the normalization is changed, see \cite{martin2}. Also,
it can be seen that the equation may be still valid for greater
values of the detuning (new terms can be treated as losses), and
diffraction in the linear medium can be taken into account
redefining the transverse coordinates $x$ and $y$.

\section{Homogeneous solutions and stability analysis} \label{stability}

Possible homogeneous solutions of Eq.\ \eqref{evol} were analyzed in
\cite{martin2},  where a classification in terms of the number of
saddle-node and pitchfork bifurcations was presented.  They can be found solving
\begin{eqnarray}\label{IntensETA}
  I_{in} &=& \left(1+ \left[\alpha \Theta-   \left(1-\frac{\bar{B}}{2}\right) I_+ -
 \left(1+\frac{\bar{B}}{2}\right)I_-\right]^2 \right)I_+ \nonumber \\
 (1-\phi) I_{in} &=& \left(1+ \left[\alpha \Theta- \left(1-\frac{\bar{B}}{2}\right)I_- -
 \left(1+\frac{\bar{B}}{2}\right) I_+\right]^2\right)I_-
\end{eqnarray}
 Where we have defined the
homogeneous solution intensities of the left and right circularly
polarized components as $I_\pm = |A_{s\pm}|^2$ (where $A_{s\pm}$ are
the stationary homogeneous solutions of Eq.\ \eqref{evol}), the
input intensity as $I_{in} = |A_{in+}|^2 + |A_{in-}|^2$, and the
polarization as $\phi = |A_{in+}|^2/I_{in}$ (the polarization $\phi$
is related to the ellipticity $\chi$ by $\phi = \cos^2(\chi/2)$).


For linearly polarized input field ($\phi = 1/2$), there is always a
symmetric linearly polarized solution, for which $I_+=I_-$. Also, a
pitchfork bifurcation may take place at $I_{in} = I'$ producing an
elliptically polarized asymmetric solution, where the upper and
lower branches correspond to either $I_+$ or $I_-$.  This new
solution may end  at $I_{in} = I''>I'$ (this happens if $\bar{B}
\Theta > 2\sqrt{1-\bar{B}}$ and $\bar{B}<1$ and is  exemplified  in
fig. \ref{homlin}, upper row, for $\bar{B}=0.9$) or may not end,
i.e. $I''=\infty$ (this happens  if $\bar{B}>1$ and is exemplified
in fig. \ref{homlin}, upper row, for $\bar{B}=1.5$).

Instead of symmetric and asymmetric solutions, for elliptic
polarization we  have continuous and discontinuous solutions.
Continuous solution  is present for any value of $I_{in}$, and a
discontinuous solution may appear at a given value of the input
intensity.  A polarization $\phi > 1/2$ favors the right circular
component for the continuous solution; but the discontinuous
solution behaves against intuition, since for $\phi > 1/2$ we have
that $I_- > I_+$.  The discontinuous solution suddenly starts at
$I_{in} = I'$, like in fig. \ref{stabellip}, upper row. It may
disappear at a second value $I_{in} = I''$. Depending on the
parameters, there are three possible situations: discontinuous
solution absent (for example, for $I' \rightarrow \infty$), bounded
($I'<I''$, both finite, like in \ref{stabellip}, upper row  for
$\bar{B}=1.2$), or left unbounded ($I'$ finite and $I'' \rightarrow
\infty$, fig. \ref{stabellip}, upper row, $\bar{B}=1.5$).


Some basic features of the homogeneous solutions can be analyzed by considering the evolution of the perturbations $\psi_\pm$ defined as
\begin{equation}
A_\pm = A_{s\pm} + \psi_\pm
\end{equation}
Replacing in \eqref{evol}, linearizing and taking the Fourier
transform (on transverse coordinates), we get
\begin{equation}
\frac{\partial}{\partial t} \left( \begin{array}{c} \textrm{Re}( \psi_+ + \psi_- ) \\ \textrm{Im}( \psi_+ + \psi_- ) \\ \textrm{Re}( \psi_+ - \psi_- ) \\ \textrm{Im}( \psi_+ - \psi_- ) \end{array} \right) = L \left( \begin{array}{c} \textrm{Re}( \psi_+ + \psi_- ) \\ \textrm{Im}( \psi_+ + \psi_- )
\\ \textrm{Re}( \psi_+ - \psi_- ) \\ \textrm{Im}( \psi_+ - \psi_- ) \end{array} \right)
\end{equation}
with the linear matrix $L$ given by
\begin{equation}
L = \left( \begin {array}{cccc} -1 &  \theta_k - \alpha S & 0 & \alpha D\bar{B}/2 \\ 3\alpha S- \theta_k & -1 & \alpha D (2-\bar{B}/2)  & 0 \\ 0 & \alpha D\bar{B}/2 & -1 & \theta_k - \alpha S \\ -3\alpha D & 0 & \alpha (1-\bar{B}) - \theta_k & -1 \end {array} \right)\label{matixL}
\end{equation}
where $S=I_+ + I_-$, $D=I_+ - I_-$, $\theta_k = \Theta + hk^2$ and $k$ is the wavenumber of the perturbation.  
Matrix $L$ has a similar form to the one derived in \cite{hoyuel1}, Eq.\ (13); 
one difference is that here the sign of the non linear term (called $\eta$ in \cite{hoyuel1}) does not have to be equal to the sign of the detuning.

The eigenvalues have the form
\begin{equation}
\lambda_{\pm\pm} = -1 \pm \sqrt{F_1 \pm \sqrt{F_2}}
\label{eigenvals}
\end{equation}
where $F_1$ and $F_2$ are real second order polynomials in $\theta_k$:
\begin{eqnarray}
F_1 &=& -\theta_k^2 + b_1 \theta_k + c_1 \nonumber \\
F_2 &=& a_2 \theta_k^2 + b_2\theta_k  + c_2
\label{efes}
\end{eqnarray}
with
\begin{eqnarray}
b_1 &=& S\, (3-\bar{B}/2)\nonumber\\
c_1 &=& S^2 (\bar{B}/2-2) + D^2\bar{B}/2 (1-\bar{B})\nonumber\\
a_2 &=& S^2 ( 1+\bar{B}/2 )^2 - 4\,\bar{B} D^2( 1-\bar{B}/2 ) \\
b_2 &=& -2 S^3 ( 1+\bar{B}/2)^2 + 3\bar{B} S D^2 (3-2\bar{B}+\bar{B}^2/4)  \nonumber\\
c_2 &=& (S^4 + D^4 \frac{\bar{B}^2}{4})(1+\frac{\bar{B}}{2})^2 + S^2 D^2  \bar{B}( 4 \bar{B} -5 - 5\frac{\bar{B}^2}{4})\nonumber
\end{eqnarray}

The homogeneous steady state solution becomes unstable when the real part of one of the eigenvalues becomes positive.  These instabilities are analyzed numerically in the next section, for linear and elliptically polarized input fields. In Sect.\ \ref{calcul} we present an analytical approach to exactly determine the values of the parameters for specific situations (codimension 2 and 3).

\section{Overview of instability regions }
\label{regions}

In this section we present a general picture of possible
patterns and instabilities that can
occur for different values of the parameters.   The parameters are
$\alpha$, the detuning $\Theta$, the sign of the refraction index
$h=\pm 1$, the non linear parameter $\bar{B}$, and the polarization
$\phi$.  We take $\alpha = 1$ and $\Theta=1$  (different values of $\Theta$, in the range
$|\Theta|<\sqrt{3}$ do not produce qualitatively
different results).  Both values of $h$ can be represented in the same marginal stability
diagram (note that the value of $h$ is not relevant for the shape
of the homogeneous solutions).

\subsection{Linear polarization}

\begin{figure}
\begin{center}
\includegraphics[scale=0.5]{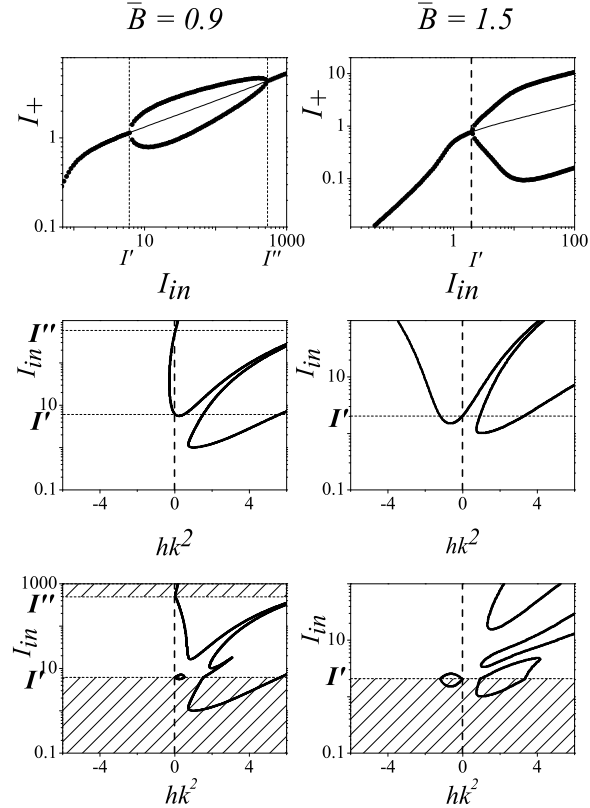}
\end{center}
\caption{Upper row: Homogeneous solutions, $I_+$ (and $I_-$) against $I_{in}$. Thin  curves correspond to regions
where solutions  become unstable under
homogeneous perturbations. Thick  curves correspond to
solutions that are stable under homogeneous perturbations. Middle and lower row:
 marginal stability curves, $I_{in}$ against $hk^2$ for symmetric (middle row) and asymmetric (lower row) solutions.  Asymmetric solutions do not exist in the striped region. In all cases, $\phi=1/2$. Left column: $\bar{B} = 0.9$; Right column: $\bar{B} = 1.5$.}
\label{homlin}
\end{figure}

From the stability analysis of the symmetric homogeneous solution we
obtain the marginal stability curves shown in Fig.\ \ref{homlin}
middle row.  In each case two unstable tongues appear, both of them
are Turing type instabilities. The lowest value of $I_{in}$ included in the  left tongue diminishes as 
$\bar{B}$ increases, while the right tongue does not depend on
$\bar{B}$ (see Fig. \ref{homlin}, middle row).  The point where the left tongue crosses the value $k=0$
corresponds to $I_{in}=I'$, i.e., it is the point where the
symmetric solution becomes unstable under homogeneous perturbations,
and the pitchfork bifurcation takes place. It is known that, for
$h=1$, for values of $I_{in}$ close and above the instability
threshold of the right tongue, an hexagonal pattern appears
\cite{firthscroggie,hoyuel1}.
  Further increase of the input intensity gives place to oscillating hexagons, quasiperiodicity and optical turbulence \cite{gomila}.


For an NRM ($h=-1$), close to the instability threshold of the left
tongue, a labyrinthic pattern is formed at short times when starting
from random initial conditions (see \cite{hoyuel1}).  For large
times, the system evolves to the homogeneous asymmetric solution.  A
competition between two regions takes place, one with $I_+>I_-$ and
the other with $I_->I_+$  (this case is illustrated in Fig.\ 4 of
Ref.\ \cite{hoyuel1}). The marginal stability curves for the
asymmetric solution are shown in Fig.\ \ref{homlin} lower row.  The
asymmetric solution is always unstable for $h=1$.  For $h=-1$, there
is a range of values of $I_{in}$ for which it can be stable.


\subsection{Elliptic polarization}
\begin{figure}[ht!]
\begin{center}
\includegraphics[scale=0.5]{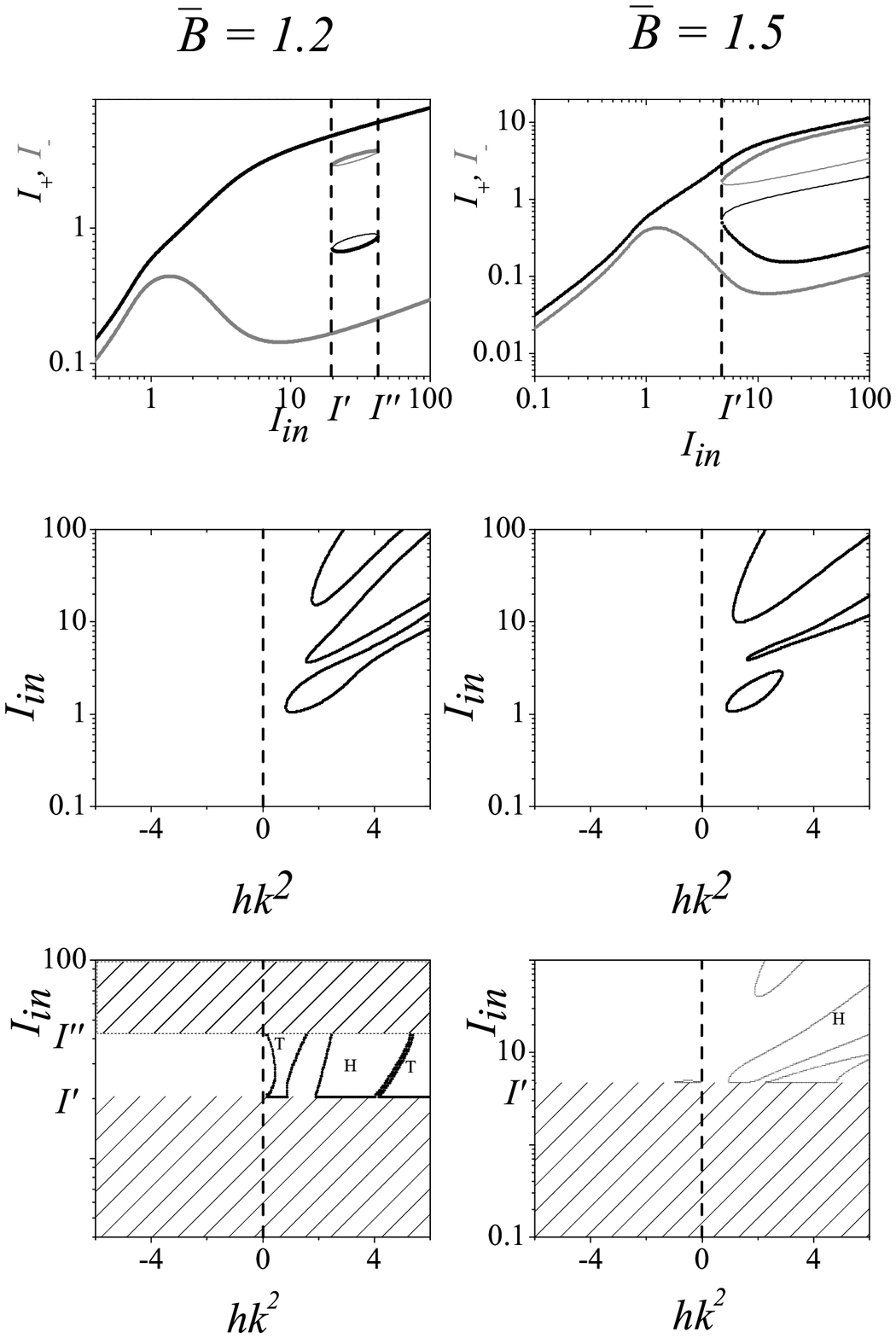}
\end{center}
\caption{Homogeneous solutions, $I_+$ and $I_-$ against $I_{in}$ (upper row) and marginal stability curves, $I_{in}$ against $hk^2$ for contiuous (middle row) and discontinuous (lower row) solutions.  Discontinuous solutions do not exist in the striped region. Labels ``H'' and ``T'' refer to Hopf or Turing instabilities. In all cases, $\phi=0.6$. Left column: $\bar{B} = 1.2$; Right column: $\bar{B} = 1.5$.}
\label{stabellip}
\end{figure}

  As the ellipticity is increased,
starting from $\phi=1/2$, for $h=-1$ the left tongue of Fig.\
\ref{homlin} middle row is transformed into a closed bounded region
whose size decreases until disappearing. For $\phi=0.6$, the
continuous solution is always stable for $h=-1$ and for any value of
$\bar{B}$, as can be seen in the marginal stability curves of Fig.\
\ref{stabellip} middle row. It can be shown that, as $\bar{B}$ is
increased, the right tongue changes its shape and is transformed
into three tongues. Two of them correspond  to Turing type
instabilities and the central tongue is related to  oscillatory in
time and usually periodic in space Hopf instability (also known as
$I_0$ type in the notation of Cross and Hohenberg
\cite{CrossyHohe}).

The stability analysis of the discontinuous solution shows that,
again, for $h=-1$ and $\phi=0.6$, it is always stable.  For $h=1$,
this solution is always unstable for some $k$.  See Fig.\
\ref{stabellip} lower row.


In general, similar plots are obtained for other values of $\phi$.
Nevertheless, a more detailed analysis of the eigenvalues in the
plane determined by $F_1$-$F_2$ in eq. (\ref{efes}) allows the
derivation of more general results and the identification of some
special cases, as explained in the next sections.

\section{Determination of instability tongues}
\label{calcul}

Since $F_1$ and $F_2$ in \eqref{eigenvals} are real quantities, it
can be shown that, if one eigenvalue becomes positive, then
$\lambda_{++}$ should be positive.  Therefore, in order to study stability, it is enough to analyze
the sign of $\lambda_{++}$.  The
analysis is simpler if, instead of describing the unstable zones in
$\theta_k$-$I_{in}$ or $hk^2$-$I_{in}$ diagrams, we first look at unstable
zones in the $F_1$-$F_2$ plane.

\begin{figure}
\begin{center}
\includegraphics[scale=0.3]{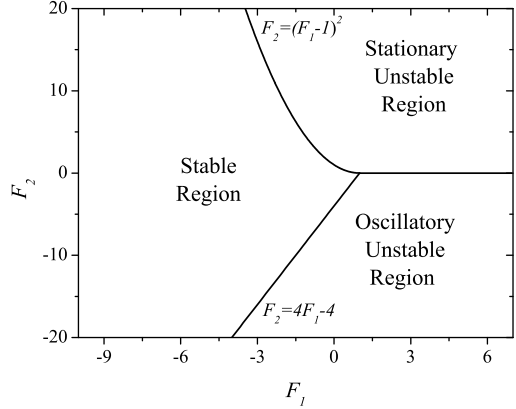}
\caption{Unstable regions in the $F_1$-$F_2$ plane.}
\label{regff}
\end{center}
\end{figure}

Using that $\lambda_{++} = -1 + \sqrt{F_1 + \sqrt{F_2}}$, we can see
that,  for $F_2<0$, points $(F_1,F_2)$ that are to the right of the
line $F_2 = 4F_1-4$ have $\textrm{Re}({\lambda_{++}})>0$ and
$\textrm{Im}({\lambda_{++}})\ne0$, therefore, it is an oscillatory
unstable region.  For $F_2>0$, points to the right of the parabola
$F_2=(1-F_1)^2$ with $F_1<1$ (that is, the left branch of the
parabola) have $\mathrm{Re}({\lambda_{++}})>0$ and
$\mathrm{Im}({\lambda_{++}})=0$, so this region is stationary
unstable.  The rest of the plane is stable, see Fig.\ \ref{regff}.

We are interested in the possible values $(F_1,F_2)$ as $\theta_k$
changes and other parameters are fixed. In a marginal instability
diagram, changing $\theta_k$ represents moving through an horizontal
line. So, if there is a value of $\theta_k$ for which $(F_1,F_2)$
falls in an unstable region of Fig. \ref{regff}, then, for that
value of $\theta_k$, in the marginal instability diagram we will be
inside an unstable tongue. Since $F_1$ and $F_2$ are second order
polynomials in $\theta_k$ \eqref{eigenvals}, the relation can be
inverted and $F_2$ can be written as  two functions of $F_1$:
$F_{2u,l}(F_1)$, which are properly defined in  Appendix A.

\begin{figure}
\begin{center}
\includegraphics[scale=0.25]{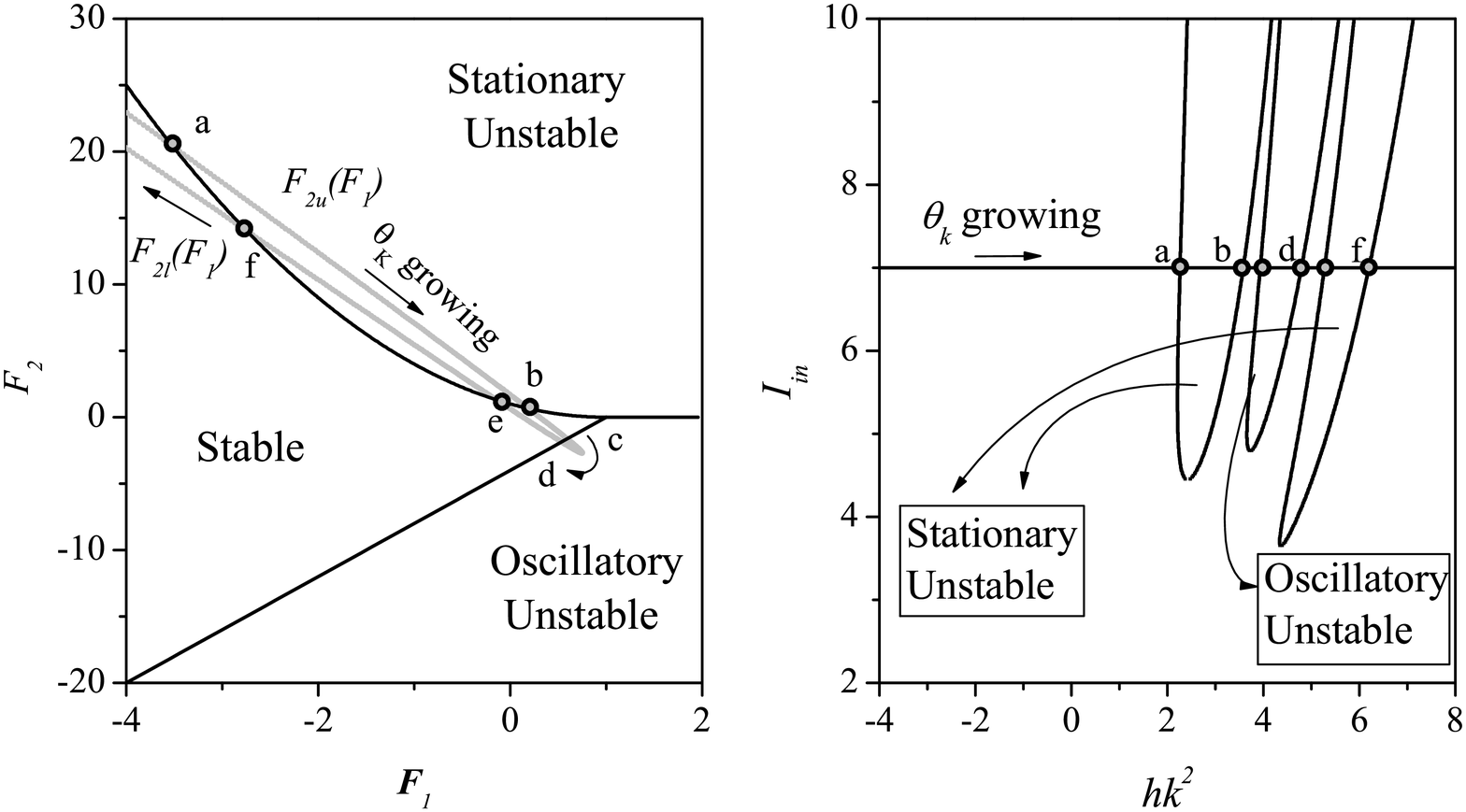}
\end{center}
\caption{Left: $F_1$-$F_2$ plane with the unstable regions and the
curves $F_{2u,l}(F_1)$; arrows indicate the direction of growing
$\theta_k$, in a situation where the maximum number of
intersections is obtained. Right: corresponding marginal stability
diagram $I_{in}$ against $hk^2$, points indicated by a, b, c, d, e
and f correspond to the intersection points of the left plot.
Parameters are: $\Theta=1$, $\bar{B}=1$, $I_{in}=7$  and
$\phi=0.8$.} \label{f1f2}
\end{figure}

An example of the relationship between marginal instability diagrams
and $F_1-F_2$ plots is graphically presented in 
Fig.\ \ref{f1f2}.  In the figure, the values of the parameters 
  were chosen in order to get the maximum
number of intersections. The left window shows values of $F_1$ and $F_2$ calculated with (\ref{efes}) as $\theta_k$ changes and all other values remain fixed. The right window shows the related points in the marginal instability diagram. The figure  shows  three unstable ranges of
$\theta_k$, the middle one is oscillatory unstable and the other two
are stationary unstable. This means that the middle tongue in the
marginal stability diagram corresponds to a Hopf instability and the
others  to Turing instabilities. This is a general behavior: it
can be shown that there can not be more than two  tongues related to
a Turing instability and one Hopf instability tongue. Also, if there
are three tongues, the middle one is the one related to Hopf
instabilities.

The intersection points of, for example, the first
tongue and a horizontal line (which represents a constant value of $I_{in}$), 
identified by `a' and `b' in the right window of Fig. \ref{f1f2}, get closer as $I_{in}$
decreases, until they merge in one point at the instability
threshold.  When we are at an instability threshold, the curve
$F_{2u,l}(F_1)$ is tangent to the border of an unstable region.  The
derivation of all these results is mathematically involved and is
sketched in Appendix A, where other
results (most of them intermediate results) are also derived.

\section{Codimension 2 and 3}

\begin{figure}[t]
\begin{center}
\includegraphics[scale=.55]{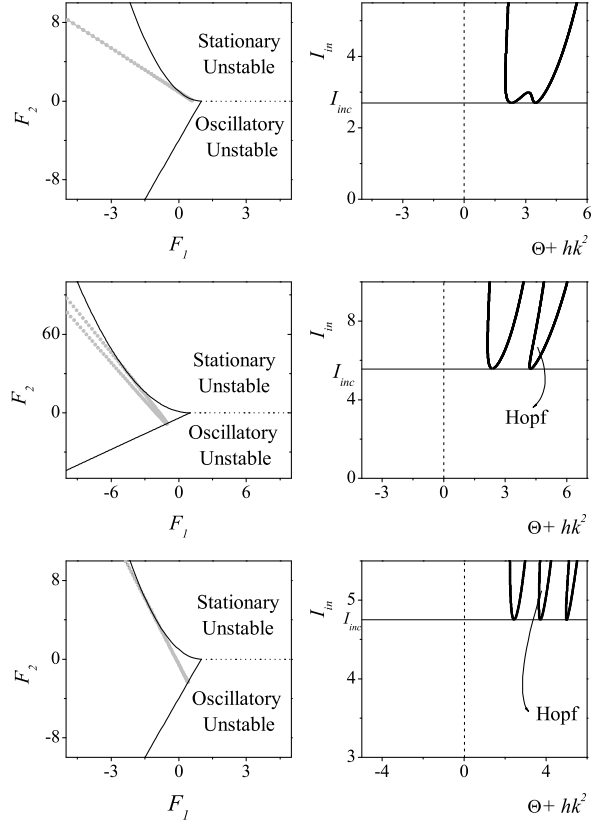}
\end{center}
\caption{Curve $F_2$ against $F_1$ (left) and marginal stability
plot (right).  From top to bottom: Turing-Turing codimension 2,
Turing-Hopf-Turing codimension 3, and Turing-Hopf codimension 2.}
\label{codim}
\end{figure}

From the previous analysis we know that we can have, at most, codimension 3 (Turinng-Hopf-Turing), i.e., 
three modes with different wavenumbers that become  unstable for the same value of $I_{in}$.
We can also have Turing-Turing codimension 2 and Turing-Hopf codimension 2. Figure \ref{codim} shows
 examples of all possible cases of codimension 2 and 3
in a $F_1-F_2$ plot and in its corresponding marginal stability
diagram.

Having in mind quite general mathematical properties and constraints given by the physical system, we are able to derive parameters for codimension 2 and 3. The values of $S$, $D$ and  $\bar{B}$ (that determine the
coefficients of $F_1$ and $F_2$ in \eqref{efes}), for which a
codimension 2 or 3 occurs, do not depend on $\Theta$. Since
$\theta_k = \Theta + hk^2$, a change in $\Theta$ produces a shift in
the marginal stability diagram. We can, in principle, take a value of $\Theta$ for
which  instability thresholds under consideration are to the
left ($h=-1$) or to the right ($h=1$) of $k=0$.

Conditions that parameters should meet in order to have codimension 2 and 3 are derived in Appendix B. We summarize the main results here.

We call $D_{TT}$ and $S_{TT}$ the values of $S$ and $D$ for a Turing-Turing codimension 2: a similar notation is used for the other cases.
For $0.848<\bar{B}<\bar{B}_c \simeq 1.028$, we have the case of
 Turing-Turing codimension 2. The values of $S$ and $D$ can be found analitically:

 \begin{equation}
S_{TT}=\frac{2\sqrt{\bar{B}(3\,\bar{B}/2-1)}}{ \bar{B}\,(1-\bar{B}/2)}
\label{STT}
\end{equation}

 \begin{equation}
D_{TT}= \pm \frac{(1+\bar{B}/2)\,S}{\sqrt{-\bar{B}^2/2 + 3\bar{B}}}.
\label{DTT}
\end{equation}

 For $\bar{B}=\bar{B}_c$, we have
codimension 3. $S_{THT}$ and $D_{THT}$ are given by $S_{THT}=
S_{TT}(\bar{B}_c)$; $D_{THT}= D_{TT}(\bar{B}_c)$.

Parameters for Turing Hopf codimenision 2 situations are harder to
determine, see  the second part of Appendix B
. After some
algebra, we find that conditions for codimension 2
Turing-Hopf situations are met only if  the roots of a given 
polinomial $P(r)$, which, once $S$ and $\bar{B}$ are fixed, is
fourth degree in an auxiliary variable $r$ (related to the difference between $F_1$ and its maximum value), has a double real and
two complex
 conjugate roots (or
  two double roots; which only happens for  $\bar{B}=\bar{B}_c$, and  corresponds to the codimension 3
  situation previously described). For every  value of $S$ and for $\bar{B}> \bar{B}_c $, we can numerically compute the roots of that polynomial.
  \begin{figure}
  \begin{center}
\includegraphics[scale=0.4]{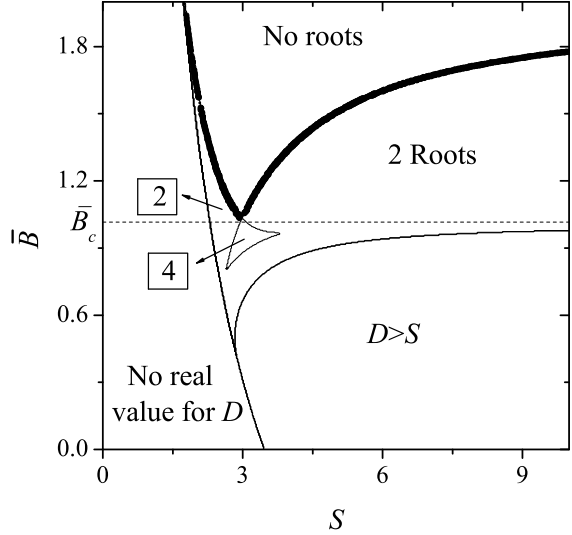}
\end{center}
\caption{Number of real roots from of $P(r)$. 
Regions where there are no possible codimension 2 situations 
(if there is no value for $D$, or if it is greater than $S$) are also shown. 
Curves in the $F_1-F_2$ diargram become tangent and thus,  Turing-Hopf codimension 2 takes place for parameters 
given by the thick line of this figure. Notice that no codimension 2 Turing-Hopf can occur for $\bar{B}<\bar{B}_c$}
\label{roots}
\end{figure}
      Figure \ref{roots} shows the number of roots in the plane $S$-$\bar{B}$. 
        The thick line represents the set of points for which codimension 2 Turing-Hopf situations take place.
        It is worth mentioning that for every allowed $\bar{B}$ there are two possible
   values of $S$: the one to the left (right) of the vertex  has a value of $\Theta+h \, k_H^2$
    greater (lower) than $\Theta+h \, k_T^2$ (where $k_{H}$ is the expected wavenumber for one of the Hopf
    instabilities, and similarly for $k_T$),  so, for $h=1$ the wavevector related to Turing instability
     is smaller (greater) than the one related to Hopf instability.  Finally, taking a point of this curve  the values of $S$,
      $\bar{B}$ and $D$ can be determined, see appendix B.

In all cases, following the derivations shown in the appendices, and
choosing a value of $\Theta$  we get the unstable wavenumbers, for instance $k_{T1}$ and $k_{T2}$  for the Turing-Turing codimension 2 situation. Conversely, we
can choose the unstable wavenumbers (for instance we can make them
fulfill a given relationship) and use that information to properly
choose $\Theta$.

Once we find $\bar{B}$, $S$ and $D$, having in mind that
$I_{\pm}=\frac{S\pm D}{2}$ and choosing a value for $\Theta$, we can
find the input intensity $I_{in}$ and polarization $\phi$ for which
codimension 2 or 3 takes place in a straightforward way (just
replacing all known values in Eq. (\ref{IntensETA}) an solving two
coupled linear equations).

\section{Numerical integration results}
Numerical integrations of  Lugiato-Lefever equations have been
extensively performed in previous reports. The novelty here is that we will exploit the results from previous
sections in order to find parameters for codimension 2 and 3
 in a straightforward way. The purpose of this section is
to  have a quick look at possible situations that may occur when
patterns tend to emerge in codimension 2 or 3.

In \cite{hoyuel1} codimension 2 Turing-Hopf situations were analyzed for the
special case $\bar{B}=1.5$. They found out that an hexagon related
to a Turing instability dominated at long times, although the Hopf
instability dominated at  short times. Also, in \cite{hoyuelTT} they
analyzed a Turing - Turing codimension 2 instability and found that
different patterns related to competition of unstable wavenumbers
might take place.

 A similar research, for a different system, was performed in
\cite{Zhabotisnky}, where pattern formation situations are analyzed
in a  Belusov-Zhabotinsky  reaction, and codimension 2 Turing-Hopf
may occur. They found out that in codimension 2 situations, patterns related to both
instabilities coexist for quite long times, but eventually  one
dominates. The exception occurs in one dimension when destabilizing
modes are  resonant (for instance, the wavelength of one instability is an
integer times the wavelength of the other instability), in that case, both
unstable modes may coexist. Similar results where found in \cite{Meixner} for a reaction-diffusion model where also chaotic situations are allowed.

In optics, two coupled Kerr-like systems (specifically, two liquid
crystal light valves) where analyzed both theoretically and
experimentally \cite{Thuring,Rubinstein,Nikolaev}.  Turing-Hopf
codimension 2 situations were reported. For some parameters,
unstable wavevectors where resonant, and a far field composed of two
octagons (whose radius where the wavevector modules  of the unstable
modes), one of them rotated $\pi/8$ degrees respect to the other,
was found \cite{Rubinstein}. Turing-Turing codimension 2 or higher
codimension was not allowed since a linearly polarized system was
studied (and, from the dynamical point of view, the system was two dimensional, i.e. instead of the
matrix in (\ref{matixL}), they had a two by two matrix).

Here, we are interested in situations where
the sum of unstable modes related to one instability may contribute
to a mode related to another instability. For instance, if one
instability is related to hexagonal patterns with some orientation and the second one has a wavelength $\sqrt{3}$ times greater,
  we expect the second one to form an hexagon $\sqrt{3}$ times larger, and rotated $\pi/6$ degrees respect to the first
  one (so that the sum of  wavevectors of the first instability should contribute to the other instability). Notice that
  parameters for which wavenumbers of the different instabilities fulfill desired relations can be found taking into
  account the calculations performed in previous sections.

Results of our numerical integrations \cite{Simul}, are the
following: At short times, all unstable wavevector coexist
(so that  $|A_{\pm}|^2$ is composed of two or three rings in the far
field), then rings of unstable wavector become thinner and the
intensity of one of them becomes much greater than the others. After
that, different situations may occur.

In some nonresonant codimension 2
 Turing-Turing situations, we found that at long times a ring of unstable modes
 (of radio $k_U$ in the far field) with wavevectors different but
  among the values of the unstable modes dominated ($k_{T1}<k_U<k_{T2}$), and the
   near field was composed of domains of ordered hexagonal patterns (for example, $A_+$ with up hexagons, and $A_-$ with down hexagons).
   For some parameters,  instead of a ring of modes, in the far field an hexagon (with $k_{T1}<k_U<k_{T2}$) was formed,
   and an hexagonal pattern arose, with a unique orientation, in the whole near field.

\begin{figure}[ht!]
\begin{center}
{\large $\mathbf{|A_+|^2}$\qquad  \qquad\qquad
$\mathbf{|A_-|^2}$}
\\
\vspace*{0.25cm}
\includegraphics[scale=0.45]{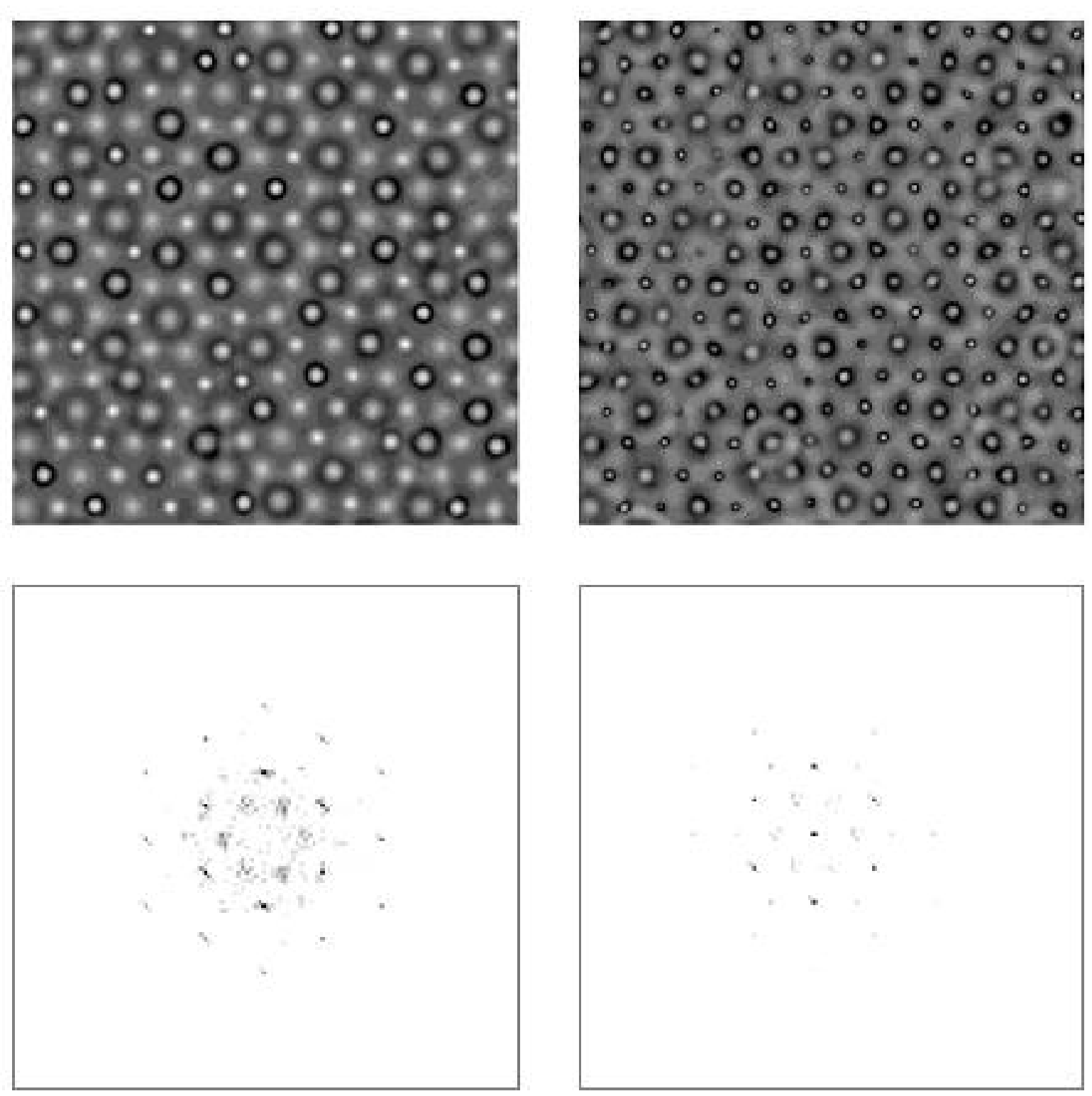} \hspace{0.5cm}  \\
\end{center}
\vspace{-0.5cm} \caption{ Left column: $|A_+|^2$, right column: $|A_-|^2$, up: near field, down: far field,  at  t=220, for the case of a resonant Turing Turing codimension 2 where initial conditions have an hexagonal modulation. The same results are found at larger times.  Parameters:
 $h=1$, $\bar{B}=0.94$, $\Theta=1.499$, $\phi=0.886$
and $I_{in}= 2.799$, $t=220$.  }
\label{cod2IC}
\begin{center}
\hspace{-1.5cm}\includegraphics[scale=0.90]{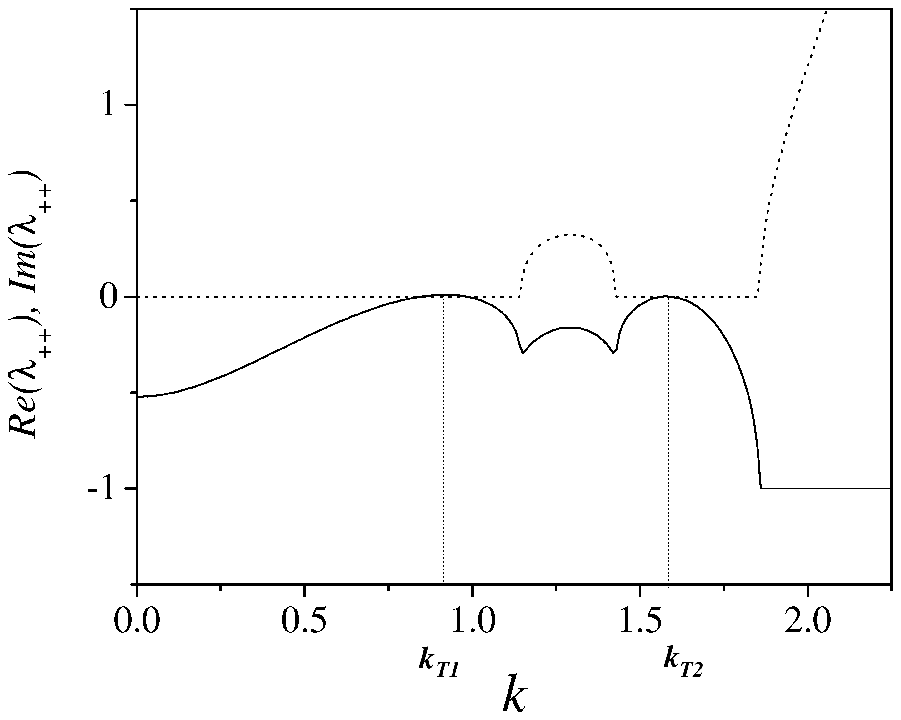}
\end{center}

\caption{Real (solid line) and imaginary (dotted line) parts of $\lambda_{++}$ as a function of $k$, for the parameters of the previous figure.
 Vertical lines show unstable wavenumbers ($k_{T1}$ and $k_{T2}$), related to steady perturbations.
 The ratio among them is $\sqrt{3}$. They closely match the numerical integration results of previous figure. } \label{fig9}
\end{figure}

   Taking an adequate value of $\Theta$,
   it is possible to make unstable wavenumbers fulfill the desired
ratio. Simulations with $k_{T1}=k_{T2}/\sqrt{3}$ were performed. For some
parameters, $k_{T2}$ dominated and a regular dodecagon took place in
the far field. Finally, putting an initial condition that was the
steady solution plus an hexagonal pattern related to the smallest
wavenumber, it could be seen that both unstable wavenumbers grew,
forming an organized hexagonal structure in the near and in the far
field. The same result was found even when the input intensity was
slightly lower than critical intensity. See Fig. \ref{cod2IC}, where, for $|A_+|^2$, the coexistence of two unstable
 wavenumbers can be found even in the near field. The stability analysis for this case is shown in Fig 
\ref{fig9}: it can be seen that there are 2 unstable wavenumbers, i.e. two values of $k$ for which $Re(\lambda_{++})$ is not negative.

\begin{figure}[ht!]
\begin{center}
\includegraphics[width=10cm]{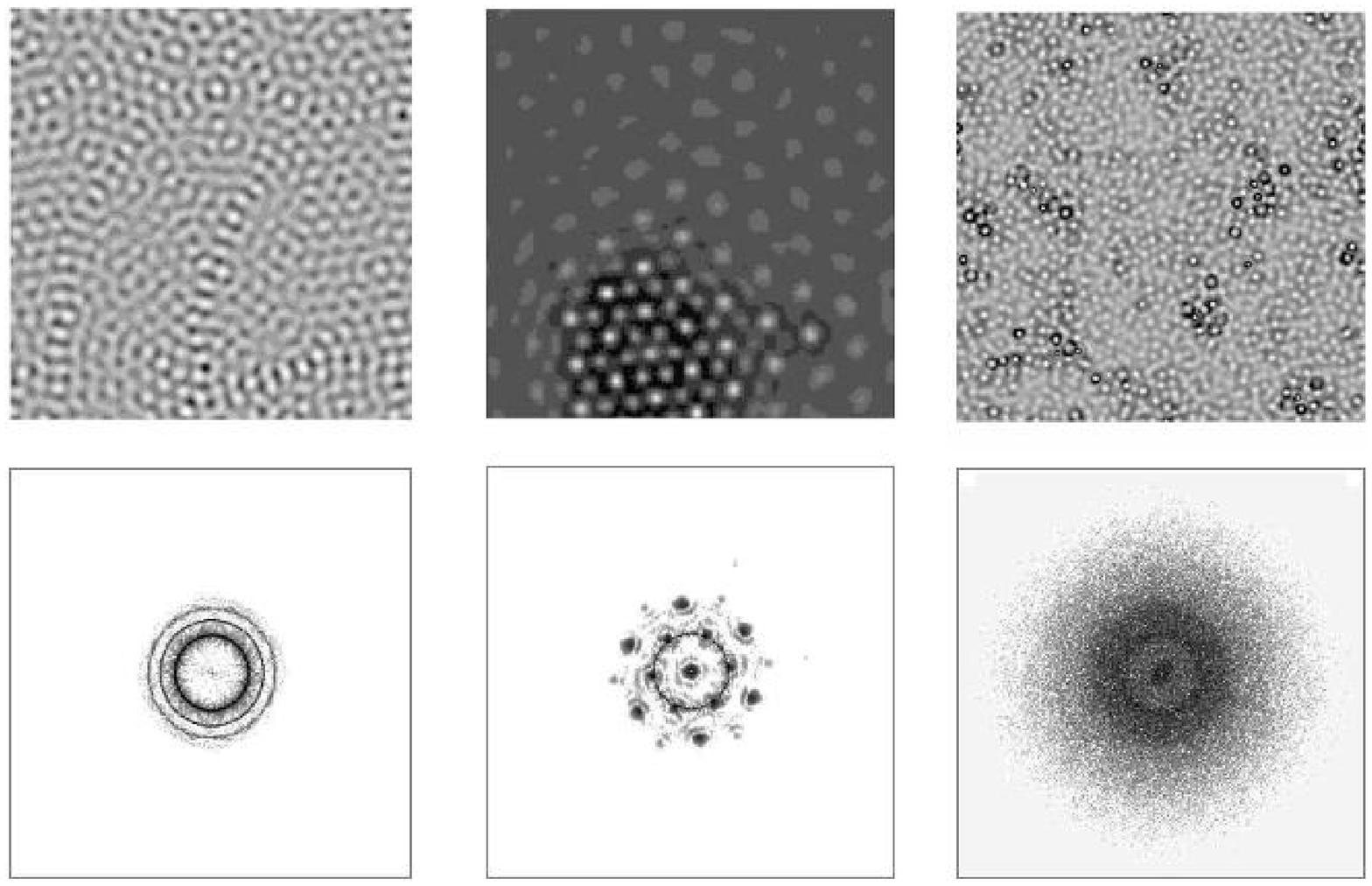} 
\end{center}
\caption{ $|A_+|^2$ in the near (up) and far field (down) at
different times for the case of resonant codimension 3. From left to
right, $t=333$ (all unstable wavenumbers are enabled), $t=482$ (an
hexagonal structure in the far field) and $t=570$. The $A_-$
component has a similar behavior. Parameters for this numerical
integration are: $h=1$, $\bar{B}=\bar{B}_c$, $\Theta=1.15$,
$\phi=0.78$ and $I_{in}= 4.26$.}
 \label{cod3}
\begin{center}
\hspace{-1.5cm}\includegraphics[scale=0.90]{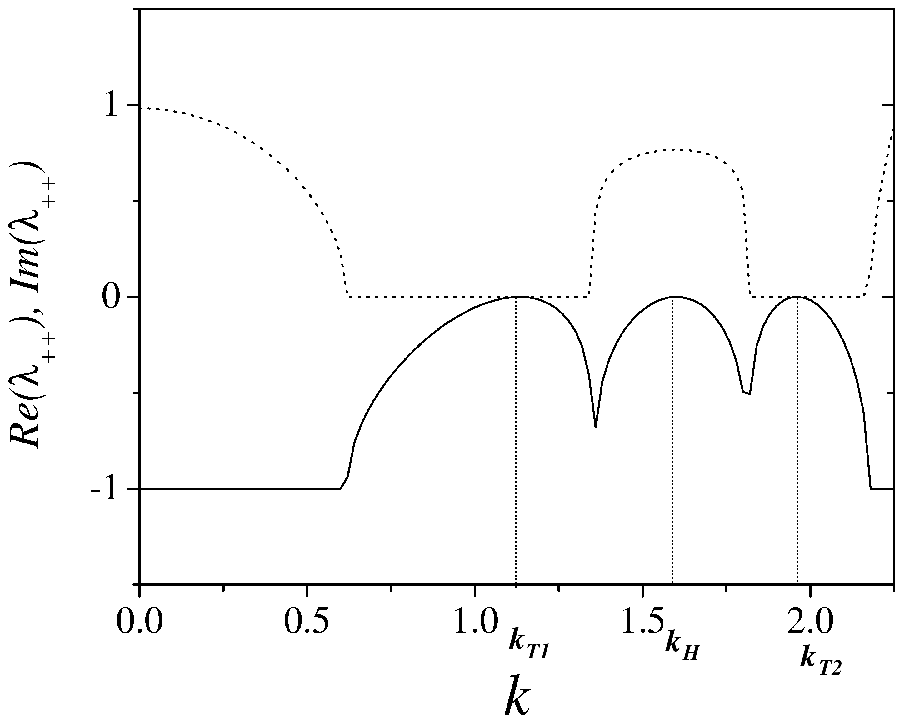}
\end{center}

\caption{Real (solid line) and imaginary (dotted line) parts of $\lambda_{++}$ as a function of $k$, 
for the parameters of the previous figure. Vertical lines show the values of the  unstable wavenumbers. 
two of them ($k_{T1}$ and $k_{T2}$) are related to steady perturbations,
 and the ratios among them is $\sqrt{3}$. The other one ($k_{H}$) is intermediate between them
  and is related to oscillatory instabilities ($Im(\lambda_{++}) \neq 0$).
 $k_{T1}$,  $k_{H}$ and $k_{T2}$ closely match the unstable wavenumbers in the previous figure. \label{fig11}}
\end{figure}

Codimension 3 situations were also analyzed.  Parameters were chosen
so that the ratio between the greatest Turing wavelength and the
smallest one was $\sqrt{3}$ ($k_{T1}<k_H<k_{T2}=\sqrt{3}k_{T1}$). At
the initial stage, we found that all unstable wavenumbers where activated (see the
left plot on fig. \ref{cod3}), forming three concentric
circumferences in the far field. The smallest Turing wavenumber
grew faster, and turned into an hexagon. After that, another hexagon,
related to the greatest Turing wavenumber also appeared, see Fig.
\ref{cod3} middle plot. At odds with fig. \ref{cod2IC}, regions where instabilities with different 
wavenumbers dominate are spatially separate in the near field. Finally, a crown of modes got destabilized
(right plot on  fig. \ref{cod3}).  The stability analysis for this case is shown in Fig \ref{fig11}. 
Notice that the intermediate unstable wavenumber is related to Hopf instability, i.e. it has  $Im(\lambda_{++}) \neq 0$.

In  another example of resonant Turing-Hopf-Turing Codimension 3, in
which $k_{T2}=2k_{T1}$, a similar final situation was observed.
For nonresonant cases, there were found steady situations similar to
fig. \ref{cod3}, middle graph, but where one Turing wavevector
(making an hexagon or a ring in the far field) prevailed for $A_+$
but the other prevailed for $A_-$.

For Turing Hopf codimension 2, situations similar to Turing Turing
codimension 2 were found, both in resonant and nonresonant
situations. Also, square patterns took place in some numerical
integrations.

\section{Conclusions}
\label{concl}

Taking into account the polarization degree of freedom of light, and having in mind that $\bar{B}$, which measures the nonlinear coupling
between different polarizations, could take a broad range of values,
we presented a study of instabilities and patterns that might show
up in a cavity filled with a Kerr-like nonlinear material with positive or negative refractive index.

A method for finding codimension 2 and 3 situations (where
wavenumbers of different modulus might destabilize) was shown. It can be found that for $0.848<\bar{B}<\bar{B}_c$ Turing-Turing codimension 2 may occur; for $B=\bar{B}_c$, Turing-Hopf-Turing codimension 3 may take place, and for  $\bar{B}>\bar{B}_c$ there may be situations for Turing-Hopf codimension 2.  Fixing only the value of $\bar{B}$, the
method  allows us to find all other values
 of the parameters for codimension 2 or 3. It
allows also to see that, for a given intensity, there cannot be more
that three instability regions in a marginal instability plot (one of
which has to be related to a Hopf instability), and that codimension   higher than 3
 cannot occur.

Since the method allows us to know some parameters with any degree
of precision   (instead of performing a numerical search and
changing the parameters until such situation shows up), and choose
others at will, it is a useful tool in the study of codimension 2 or
3 on the model. Specifically, resonant situations, where the ratios
between unstable wavevectors are chosen, can be found. Also, it
might be useful for the underestanding of pattern formation in other
systems as long as the linear stability analysis presents
eigenvalues with the form of eqs. (\ref{eigenvals}) and
(\ref{efes}).

Numerical integration results show some new situations of pattern coexistence and competition.

\section*{Acknowledgments}
This work was partially supported by Consejo Nacional de
Investigaciones Cient\'{\i}ficas y T\'{e}cnicas (CONICET, Argentina, PIP 0041
2010-2012).

\section*{Appendix A: Properties of instability tongues}
\label{apend1}

 We consider that the possible values
$(F_1,F_2)$ are those given by these polynomials with the free
parameter $\theta_k$, and with fixed coefficients, i.e., we consider
fixed values of $S$ and $D$, that correspond to a fixed value of the
input intensity $I_{in}$  for a determined homogeneous solution.
From \eqref{efes} we see that $F_1$ takes a maximum value given by
$F_{1M} = b_1^2/4 + c_1$.  By definition, $S\ge D$, so it can be
shown that $a_2 \ge 0$, and that $F_1 + \sqrt{F_2} \rightarrow
-\infty$ for $|\theta_k| \rightarrow \infty$.  This means that, for
large $|\theta_k|$, $\textrm{Re}(\lambda_{++})=-1$. Therefore, the
range of unstable wavenumbers is bounded. The case $a_2=0$ occurs
only if $S=D$, that is, for pure circular polarization, but this
case can be related to the pure linear polarization case, as has
been done, for instance in \cite{hoyuel1}.

 From \eqref{efes}, we can obtain
two solutions for $\theta_k$ as a function of $F_1$. Using these
solutions in the equation for $F_2$, we obtain:
\begin{equation}\tag{A.1} 
F_{2u,l}(F_1)= (a_2\,b_1^2/2+b_2\,b_1/2+c_2+a_2\,c_1)-a_2 \,F_1 \pm
|X|\, r, \label{f2ul}
\end{equation}
where $X=a_2 b_1+b_2$ and $r= \sqrt{ F_{1M}-\,F_1}\geq 0$; indices `{\textit{u}}' and `\textit{l}' stand for upper and lower curves respectively.

The difference between the upper and lower curves is
$F_{2u}(F_1)-F_{2l}(F_1)= |X|r$. So,  we have two values of $F_2$
for each $F_1$ as long as $F_1 < F_{1M}$ and $X\ne 0$.  For $X=0$,
the curves are two overlapping rays that start from $F_1 = F_{1M}$.
We can also see that $\frac{\partial^2 F_{2u,l}}{\partial F_1^2}=\mp
2|X|/r$, so $F_{2u}$ ($F_{2l}$) has negative (positive) curvature.

To obtain the instability points, we have to look at the intersection of $F_{2u,l}(F_1)$ with the unstable regions of Fig.\ \ref{regff}.  Eq.\ \eqref{f2ul} can be rewritten as:
\begin{equation}\tag{A.2}
(F_2 - d + a_2 F_1)^2 = X^2 (F_{1M}-F_1), \label{f2ul2}
\end{equation}
where $d=a_2\,b_1^2/2+b_2\,b_1/2+c_2+a_2\,c_1$ (the same equation holds for $F_{2u}$ and $F_{2l}$). To obtain the intersection with the stationary unstable region, we replace $F_2$ by $(1-F_1)^2$ in \eqref{f2ul2}.  We get a 4th order polynomial, so there are at most 4 solutions for $F_1$.  The intersection with the oscillatory unstable region is obtained replacing $F_2$ by $4F_1-4$ in \eqref{f2ul2}; this gives two solutions for $F_1$.  To obtain the maximum number of unstable ranges of values of $\theta_k$ we assume that half of the previously mentioned solutions of $F_1$ correspond to a cross from stable to unstable region (as $\theta_k$ is increased) and the other half to a cross from unstable to stable region. So, we have, at most, three unstable ranges of $\theta_k$ that correspond to three tongues in the marginal stability diagrams.
\section*{Appendix B: Derivation of parameters valid for codimension 2 and 3}\label{apen2}
For Turing-Turing codimension 2, we need  both $F_{2u}$ and
$F_{2l}$ to be tangent to the border of the stationary unstable region,
and this happens only if $X=0$; in this case $F_{2u}$ and $F_{2l}$
are straight lines that overlap.   To have codimension 3
(Turing-Hopf-Turing) there is a further condition: the end point of
the rays, $(F_{1M}, F_2(F_{1M}))$, must be on the border of the
oscillatory unstable region given by $F_2 = 4F_1 -4$.  The case of
Turing-Hopf codimension 2 occurs when $F_2$ against $F_1$ is tangent
to the borders of the stationary and oscillatory unstable regions,
and $X \ne 0$. See Fig. \ref{codim}.

\subsection*{Turing-Turing codimension 2 and Turing-Hopf-Turing codimension 3}

 From the condition $X=a_2 b_1+b_2=0$ it is straightforward to obtain an expression for $D_{TT}(\bar{B})$, (\ref{DTT}).
From Eq.\ \eqref{f2ul2} we have that $F_2= d - a_2 F_1$, and the
intersection with the border of the stationary unstable region,
given by $F_2 = (1-F_1)^2$, gives a 2nd order polynomial in $F_1$.
To have the line tangent to the parabola, the discriminant of the
polynomial should be zero. From this conditions we get the critical
value $F_{1c}=1-a_2/2$ from which, using Eq.\ \eqref{efes}, we get
the two critical values of $\theta_k$.  From the zero discriminant
and Eq.\ \eqref{DTT}, we get an expression for $S_{TT}(\bar{B})$,
(\ref{STT}).

Then, for a given value of $\bar{B}$, there is a unique value of $S_{TT}$ and $|D_{TT}|$ where we can find Turing-Turing codimension 2.
 There are some restrictions on the possible values of $\bar{B}$.  First, in order to have $S_{TT}$ real, we have that $\bar{B} \ge 2/3$,
 but there is a more restrictive condition. We need that $F_{1c} \le F_{1M}$ in order to  have a solution tangent to the unstable border
 that actually touches it.  
 It can be shown
that the condition $\bar{B} > 0.848$ should be satisfied. Second,
the Hopf instability should appear for greater values of $I_{in}$
than the Turing-Turing instability.



We define the distance between $F_2(F_{1M})$ and the border of the oscillatory unstable region as $Z=F_2(F_{1M})-(4 \,F_{1M}-4)$
. For Turing-Turing codimension 2 we need $Z>0$, and for codimension 3, we have that $Z=0$ since the point $(F_{1M},F_2(F_{1M}))$ should be on the oscillatory unstable border.  Since $S_{TT}$ and $D_{TT}$ are functions of $\bar{B}$ [see Eqs.\ \eqref{STT} and \eqref{DTT}], we can obtain $Z$ as a function of $\bar{B}$ only.  It can be shown that the only zero of Z occurs for $\bar{B}_c = 1.028$. Then, using the value $\bar{B}=\bar{B}_c$, we can obtain the parameters $S_{THT} = S_{TT}(\bar{B}_c)$ and $D_{THT} = D_{TT}(\bar{B}_c)$ for codimension 3. For $\bar{B}<\bar{B}_c$, $Z>0$, so that Turing-Turing codimension 2 is allowed. For $\bar{B}>\bar{B}_c$,
$Z<0$: we still have the two stationary instabilities that occur
simultaneously for a given value of $I_{in}$, but this is not
Turing-Turing codimension 2 since the oscillatory instability
appears for a smaller value of $I_{in}$ (As we will see below, in
that region there are Turing-Hopf instabilities).

Expressions for $\Theta+ h k_{T1}^2$ and $\Theta+h k_{T2}^2$ can be found solving $F_1=1-a_2/2$, and replacing $S$ by $S_{TT}$ and $D$ by $D_{TT}$. If there is codimension 3, the value of $\Theta+h k_H^2$ can be found from $\Theta+h k_H^2=b_1/2$. It can be shown that $2 h k_H^2= h k_{T1}^2 + h k_{T2}^2$.  Once the value of $\Theta$ is chosen, $ k_H$, $k_{T1}$ and $k_{T2}$ are fixed. Conversely, once two wavenumbers are chosen, $\Theta$ is fixed (and so is the third wavenumber, if exists).

\subsection*{Turing-Hopf codimension 2}\label{subsecTH}

For a Turing-Hopf codimension 2 we require the curve $F_2(F_1)$ to
be tangent to both borders of the unstable regions, as shown in the
lower row of Fig.\ \ref{codim}.

Let us first consider the contact point with the border of the
oscillatory unstable region, i.e., between $F_{2l}(r)$
\eqref{f2ul} and the line $F_2 = 4F_1 - 4 = 4(F_{1M} - r^2)-4$.  The
intersections are given by a second order polynomial in $r$.  We
require the intersection to be only in one point, so that the
polynomial discriminant should be zero.  From this condition, it is
possible to obtain three possible expressions for $D$ as a function
of $S$ and $\bar{B}$. We will call them $D_{1,2,3}(S,\bar{B})$.

 Now,  we consider the intersection with the stationary unstable region, i.e., between $F_{2u}(r)$
  and $F_2 = (1-F_1)^2= (1 - F_{1M} + r^2)^2$.
   We get a fourth order polynomial in $r$, which will be called $P(r)$.  We can have 0, 2 or 4 real roots, and we are interested in the cases of a
   fourfold real root, or double real and two complex conjugate roots, in order to have the function $F_{2u}$
   tangent to the unstable border.  It can be shown that the kind of roots that we are looking for are possible only for
    one of the expressions of $D$ mentioned in the previous paragraph, say $D_1(S,\bar{B})$. In fig. \ref{roots} we plot the roots of $P(r)$. Its coefficients are calculated for $S$, $\bar{B}$ and $D=D_1(S,\bar{B})$.

The value of $\Theta+h k_T^2$ can be found solving $F_1|_{\theta_k=\Theta+h k_T^2}= F_{1M}-r_0^2$, where $r_0$ is the double root of $P(r)$. The value of $\Theta+h k_H^2$ can be found in a similar way.


\section*{Bibliography}
\begin{thebibliography}{00}

\bibitem{lugiatoRev} L. A. Lugiato, M. Brambilla, A. Gatti, Optical pattern fomation, Advances in Atomic, Molecular and Optical Physics {40} (1999) 229.


\bibitem{LibroST} K. Staliunas, V.J. Sanchez-Morcillo, \emph{Transverse Patterns in Nonlinear Optical Resonators}, Springer Verlag, Springer Tracts in
Modern Physics, Vol.183, 2003.


\bibitem{hoyuel1} M. Hoyuelos, P. Colet, M. San Miguel, D. Walgraef, Polarization patterns in Kerr media, Phys. Rev. E {58} (1998) 2992.

\bibitem{Yano} V. Yannopapas, Enhancement of nonlinear susceptibilities near plasmonic metamaterials, Opt. Commun. 283 (2010) 1647.

\bibitem{SipeyBoyd} J. E. Sipe, R. W. Boyd, Nonlinear susceptibility of composite optical materials in the Maxwell Garnett model, Phys. Rev. A {46} (1992) 1614.







\bibitem{rama}  S. A. Ramakrishna, Physics of negative refractive index materials, Rep. Prog. Phys. {68} (2005) 449.




\bibitem{Merlin} R. Merlin,  Metamaterials and the Landau$-$Lifshitz permeability argument: Large permittivity begets high$-$frequency magnetism, 
PNAS  106 (2009) 1693-1698.


\bibitem{zharov} A. A. Zharov, I. V. Shadrivov, Y. S. Kivshar, Nonlinear Properties of Left-Handed Metamaterials, Phys. Rev. Lett. {91} (2003) 037401.

\bibitem{Negdifrac}  R. Morandotti, H. S. Eisenberg,  Y. Silberberg, M. Sorel, J. S. Aitchison, Self-Focusing and Defocusing in Waveguide 
Arrays, Phys. Rev. Lett. {86} (2001) 3296.

\bibitem{paperST2} K. Staliunas, R. Herrero, Nondiffractive propagation of light in photonic crystals, Phys. Rev. E {73}  ͑(2006͒) 016601.

\bibitem{paperST}  Staliunas, O. Egorov, Y. S. Kivshar, F. Lederer, Bloch Cavity Solitons in Nonlinear Resonators with Intracavity Photonic
Crystals Phys. Rev. Lett. 101, 153903 (2008)


\bibitem{martin2} D. A. Martin, M. Hoyuelos, Homogeneous solutions for elliptically polarized light in a cavity containing materials
with electric and magnetic nonlinearities, Phys. Rev. A {82} (2010) 033841.



\bibitem{Tassin} P. Tassin, G. Van der Sande, I. Veretennicoff, M.
Tlidi, P. Kockaert, Analytical model for the optical propagation in a nonlinear left-handed material, Proc. SPIE 5955 (2005) 59550X.





\bibitem{nosotros} D. A. Martin, M. Hoyuelos, Cavity equations for a positive- or negative-refraction-index material
with electric and magnetic nonlinearities, Phys. Rev. E {80} (2009) 056601.



\bibitem{lugiato} L. A. Lugiato, R. Lefever, Spatial Dissipative Structures in Passive Optical Systems, Phys. Rev. Lett. {58} (1987) 2209.

\bibitem{boyd}  R. W. Boyd, \emph{Nonlinear Optics}, third ed., Academic, New York,
2007.


\bibitem{Bugin}  J. Bugin, C. Gillon, P. Langot, Femtosecond investigation of the non-instantaneous third-order nonlinear suceptibility in liquids and glasses, Appl. Phys. Lett. {87} (2005) 211916.





\bibitem{firthscroggie}  A. J. Scroggie, W. J. Firth, G. S. McDonald,  Pattern formation in a passive Kerr cavity, Chaos, Solitons \& Fractals {4}  (1994) 1323.

\bibitem{gomila} D. Gomila, P. Colet, Transition from hexagons to optical turbulence, Phys. Rev. A {68} (2003)
011801.

\bibitem{CrossyHohe} M. C. Cross,  P. C. Hohenberg, Pattern Formation Out of Equilibrium, Rev. Mod. Phys. {65} (1993) 851.

\bibitem{hoyuelTT} M. Hoyuelos, D. Walgraef, P. Colet, M. San Miguel, Patterns arising from the interaction between scalar and vectorial instabilities in two-photon resonant Kerr cavities, Phys. Rev. E {65} (2002) 046620.

\bibitem{Zhabotisnky} L. Yang, M. Dolnik, A. M. Zhabotinsky, I. R. Epstein, Pattern formation arising from interactions between Turing and wave instabilities, J. Chem. Phys.{117} (2002) 7259.

\bibitem{Meixner} M.  Meixner,  S.  Bose,  E.  Scholl,  Analysis  of complex  and  chaotic  patterns  near  a codimension-2
Turing-Hopf  point in  a reaction-diffusion  model, Physica D  109 (1997) 128-138.

\bibitem{Thuring} B. Th\"{u}ring, A. Schreider, M.Kreuzer, T. Tschudi, Spatio-temporal  dynamics  due to competing spatial instabilities  in a
coupled LCLV feedback system, Physica D 96 (1996) 282-290.

\bibitem{Rubinstein} B. Y. Rubinstein, L. M. Pismen, Resonant patterns in a two-component optical system
with 2-D feedback, Opt. Commun. 145 (1998) 159-165.

\bibitem{Nikolaev} I.P. Nikolaev, A.V. Larichev, E.V. Degtiarev, V. Wataghin, An optical feedback nonlinear system with a Takens-Bogdanov
point: experimental investigation, Physica D 144 (2000) 221-229.



\bibitem{Simul} Numerical integrations were performed in a 256 by 256 square matrix, with periodic boundary conditions. Evolution was calculated with a fourth order Runge Kutta method. Time step was $dt=0.0004$.


%
%
%
%
%
%
%


\end{thebibliography}
\end{document}